\documentclass[12pt, draftclsnofoot, onecolumn]{IEEEtran}
\usepackage{multirow}
\usepackage{graphicx}
\ifCLASSOPTIONcompsoc
    \usepackage[caption=false, font=normalsize, labelfont=sf, textfont=sf]{subfig}
\else
\usepackage[caption=false, font=footnotesize]{subfig}
\fi
\usepackage{dcolumn}
\usepackage{url}

\usepackage{cite}

\ifCLASSINFOpdf
\else
\fi
\usepackage{amsmath}

\usepackage{bm}
\usepackage{graphicx}
\usepackage{caption}
\usepackage[dvipsnames,table,xcdraw]{xcolor}
\usepackage{amssymb}
\usepackage{amsmath}

\begin{document}
%
\title{Frequency Plan Design for Multibeam Satellite Constellations Using Linear Programming}
%
%
%

\author{Juan~Jose~Garau-Luis, 
        Sergi~Aliaga, 
        Guillem~Casadesus, 
        Nils~Pachler,
        Edward~Crawley, 
        and~Bruce~Cameron%
\thanks{J. Garau-Luis, G. Casadesus, N. Pachler, E. Crawley, and B. Cameron are with the Aeronautics and Astronautics department, Massachusetts Institute of Technology, Cambridge, MA, USA. S. Aliaga is with the Electrical Engineering and Computer Science department, Northeastern University, Cambridge, MA, USA.}
\thanks{Corresponding author: Juan Jose Garau-Luis (\textit{garau@mit.edu})}%
\thanks{This work has been submitted to the IEEE for possible publication.  Copyright may be transferred without notice, after which this version may no longer be accessible.}}

\maketitle

\begin{abstract}
Upcoming large satellite constellations and the advent of tighter steerable beams will offer unprecedented flexibility. This new flexibility will require resource management strategies to be operated in high-dimensional and dynamic environments, as existing satellite operators are unaccustomed to operational flexibility and automation.
Frequency assignment policies have the potential to drive constellations' performance in this new context, and are no exception to real-time and scalability requirements. The majority of frequency assignment methods proposed in the literature fail to fulfill these two requirements, or are unable to meet them without falling short on bandwidth and/or power efficiency. In this paper we propose a new frequency assignment method designed to prioritize operational requirements. We present an algorithm based on Integer Linear Programming (ILP) that is able to fully define a frequency plan while respecting key system constraints such as handovers and interference. We are able to encode operators' goals such as bandwidth maximization or power reduction and produce optimal or quasi-optimal plans according to such objectives. In our experiments, we find our method is able to allocate at least 50\% more bandwidth and reduce power consumption by 40\% compared to previous operational benchmarks. The performance advantage of our method compared to previous solutions increases with the dimensionality of the constellation; in an experiment with a 5,000-beam MEO constellation we find that we can allocate three times more bandwidth.
\end{abstract}

\begin{IEEEkeywords}
Satellite communications, frequency assignment, multibeam constellations, resource management
\end{IEEEkeywords}

%
\IEEEpeerreviewmaketitle





\section{Introduction}
%
%
%
%

\IEEEPARstart{A}{lthough} satellite operators have been in the market for many years, their operations still tend to be relatevely static---in time. This is especially the case for spectrum allocation policies. When frequency plans do vary, the changes are generally manually-operated, in an effort to find margin. This process entails dealing with a complex optimization problem that has been long studied. However, upcoming changes in the satellite communications industry and the need for automation pose additional challenges to spectrum optimization which previous methods might not be able to address; new algorithmic approaches are required instead.



\subsection{Motivation}

Three trends define the new satellite communications landscape. First, \emph{new types of users} are coming into the market. Spaceborne data services are expected to substantially grow in the coming years \cite{NorthernSkyResearch2019}, and mobile users will constitute a significant fraction of this new demand \cite{NorthernSkyResearch2019a}. In the past, operators could safely assume fixed terminal locations; moving forward, frequency assignment policies must account for location changes over time, which at the technical level involves considering time-dependant constraints. There is a need for effective mechanisms against highly dynamic and time-dependent environments.

The remaining two trends are a consequence of technological improvements. On the one hand, the introduction of \emph{highly flexible payloads} 
is constituting a market disruption in terms of spectrum management strategies. Previously, frequency plans were generally fixed, but in the next generation of communication satellites, operators will be able to reconfigure frequency plans in orbit, intermittently, and on a beam level. Exploiting this flexibility, as well as other mechanisms such as frequency reuse, is key to reacting to demand changes and achieving efficient spectrum usage.

On the other hand, the \emph{dimensionality} of some of the upcoming constellations constitutes an additional level of complexity. Examples include SpaceX's 4,408-satellite LEO constellation with up to 32 beams per satellite \cite{DelPortillo2019,Pachler2021} and SES's O3b mPOWER MEO constellation, consisting of 11 satellites able to power thousands of beams each \cite{SESSA2021}. 
Any future frequency management strategy needs to account for scalability, which challenges the adequacy of some spectrum optimization approaches considered in the past.

This new context calls for revisiting the strategies and algorithms previously utilized to make frequency assignment decisions. This problem is not only NP-hard and highly constrained \cite{Mizuike1989}, but now it also demands algorithmic solutions that are fast and scale well in order to be successfully operated in upcoming high-dimensional and dynamic environments.

\subsection{Literature Review}





Previous studies focus on carrier assignment (i.e., which central frequency), bandwidth allocation (i.e., how much bandwidth), or both. Reference \cite{Camino2014} addresses the carrier assignment problem, specifically targeting the recoloring problem by means of local search and the Simulated Annealing algorithm, after an initial greedy combinatorial optimization approach. With the scaling computational burden of classic optimization approaches in mind, other authors propose Artificial Intelligence (AI) algorithms as an alternative to try to meet the new requirements of the future scenarios. In \cite{Hu2018}, a Deep Reinforcement Learning (DRL) model is proposed to solve the Dynamic Channel Allocation (DCA) problem, and it is shown to closely match the performance of state-of-the-art DCA algorithms. While the online operation of such models consists of simple forward passes of a neural network -- substantially faster than other methods -- the robustness and stability of DRL models is still disputed in many real world scenarios \cite{Dulac-Arnold2021}.

Regarding bandwidth allocation studies, in \cite{Park2012}, a binary search-based method is proposed as a fairer alternative to the water-filling approach considered in \cite{Choi2005}. A later study \cite{HengWang2013} also uses the same fairness-centered optimization objective but allocates bandwidth using convex optimization instead. However, the nature of future satellite constellations is not fully reflected in their analyses since both studies test their methods on single-satellite and single-gateway use cases with no more than 20 beams. In the AI domain, in \cite{paris19} and \cite{Ferreira2018}, Genetic Algorithms and DRL are, respectively, used to allocate bandwidth in single-satellite systems. Both studies consider simultaneously optimizing other variables such as power or the roll-off factor, respectively, but the solution quality and robustness of these AI methods are not tested for larger systems. 

Both classic optimization and AI approaches are also considered for the combined carrier assignment and bandwidth allocation problem. In \cite{Salcedo-Sanz2005} a neural network combined with a genetic algorithm is used to address the problem through an interference minimization lens; it is tested on multiple scenarios, with a maximum of 36 beams and 128 bandwidth channels. The same problem, with an extension to also optimize for power, is addressed in \cite{Cocco2018} using a capacity-oriented objective function and the Simulated Annealing algorithm. Tests on a single satellite with 200 beams and 16 bandwidth channels show that this method reduces both the unmet capacity and the excess capacity in comparison to traditional approaches based on conventional payloads. While these studies focus on optimizing bandwidth and carrier simultaneously, further considerations such as frequency reuse mechanisms or dynamic environments are out of their scope. 

Two works do account for reuse mechanisms and prove to be substantially faster than other approaches. First, the DRL-based method introduced in \cite{Luis2021} uses an agent to simultaneously make carrier assignments, bandwidth allocation, and frequency reuse selection. Despite its ability to scale, this algorithm does not always satisfy the complete set of constraints. Second, the constraint satisfaction algorithm presented in \cite{PachlerdelaOsa2020} uses a greedy assignment strategy to produce valid frequency plans in short amounts of time. However, since beams are assigned one at a time without an objective function, its usefulness is unclear when the operator wishes to prioritize a specific metric (e.g., bandwidth maximization).

\subsection{Paper Objectives}
Given the new industry needs and the flexibility and dimensionality of new systems, an efficient spectrum management algorithm for multibeam satellite constellations must be able to simultaneously address the carrier assignment and the bandwidth allocation problems for constellations with up to thousands of satellites or beams. In addition, new frequency reuse mechanisms should be taken into account as key efficiency drivers, and an algorithm with a small computing burden is preferred given the dynamic behavior of the users. All the studies presented fail to address this complete set of requirements. Furthermore, none consider optimizing for multiple satellites simultaneously, and frequency reuse mechanisms are rarely included as an additional optimization dimension.

To try to close this research gap, we propose a frequency plan design algorithm based on Integer Linear Programming (ILP) that optimizes for carrier assignment, bandwidth allocation, and frequency reuse on the beam level in multibeam satellite constellations. The algorithm produces optimal or quasi-optimal plans according to an objective function, and these plans respect interference and handover constraints. The objective function can reflect multiple goals, such as maximizing bandwidth, minimizing frequency reuse, and minimizing power. We show that our algorithm can be adapted to address multiple scenarios including non-Geostationary Orbits (NGSO), dynamic environments, and high-dimensional constellations. 

\subsection{Paper Structure}

The remainder of this paper is structured as follows:
Section \ref{sec:prob_statement} presents the frequency plan design problem, the assumptions considered and the constraints involved;
Section \ref{sec:opt_framework} introduces the optimization method based on Integer Linear Programming, together with the requirements that can be encoded in the objective function;
Section \ref{sec:results} discusses the results of applying our optimization algorithm to different scenarios; and finally
Section \ref{sec:conclusions} remarks the conclusions of the paper.


\IEEEpubidadjcol

\begin{figure*}[!t]
\centering
\includegraphics[width=.9\linewidth]{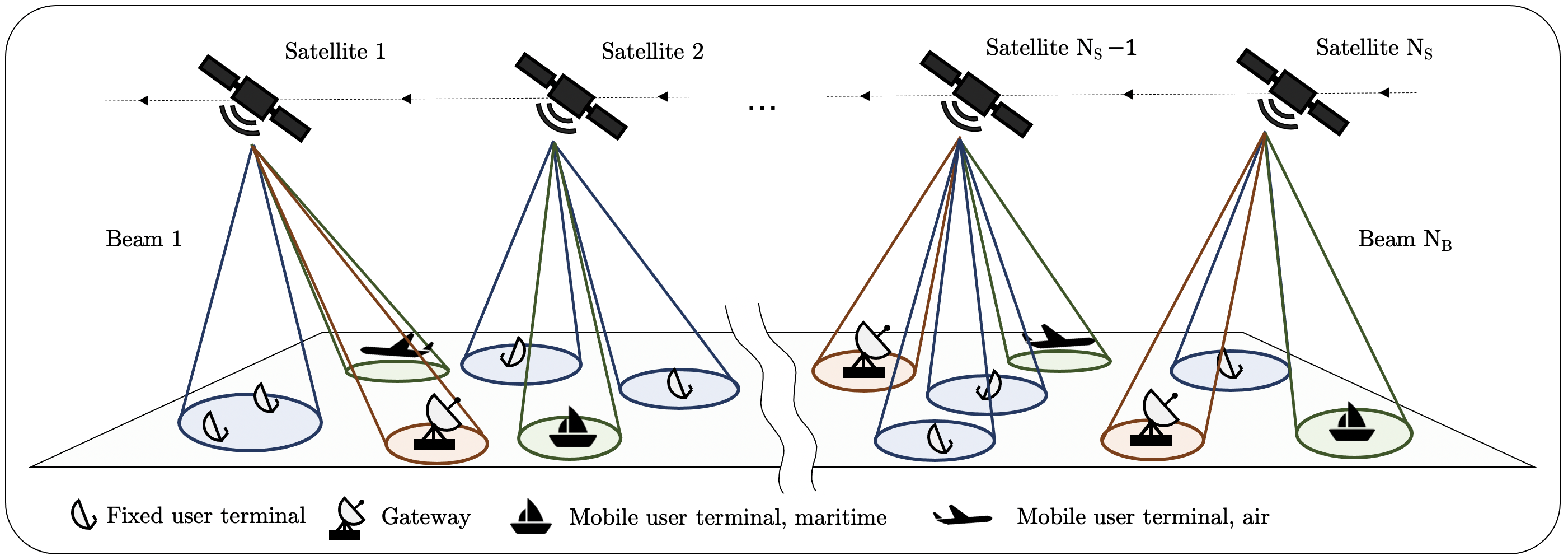}
\caption{A constellation with $N_S$ identical satellites in the same orbit and $N_B$ beams is considered. Gateways, fixed terminals, and mobile users are connected to the network.}
\label{fig:constellation}
\end{figure*}

\section{Problem Statement}
\label{sec:prob_statement}
As depicted in Figure \ref{fig:constellation}, we consider a satellite constellation with $N_S$ identical multibeam satellites, all of them in the same orbital plane. The constellation serves multiple users and gateways on Earth, which we assume have already been grouped into $N_B$ beams. Each beam constantly serves its gateway or its group of users, which might be mobile ---the beam then ``follows'' the users. At any point in time, each beam is powered from one ---and just one--- of the satellites of the constellation. If the orbit of the satellites is NGSO, then handover operations take place and the satellite powering each beam changes over time. Our goal is to design a \textit{frequency plan} for this constellation, which consists of completely defining frequency utilization for all beams.

In terms of frequency resources, we assume all satellites are allowed to use the exact same part of the spectrum, which is divided into $N_{BW}$ equal bandwidth chunks or slots. Likewise, all satellites have identical frequency reuse mechanisms: 

there are $N_{FR}$ frequency reuses available, as well as $N_P$ polarizations for each reuse. Polarizations allow to use more spectrum in a concentrated area without incuring into additional interference. For example, $N_P=2$ when using right-handed and left-handed circular polarizations (RHCP and LHCP, respectively). For each beam, the operator must decide how many bandwidth slots are assigned and which reuse group and polarization should be used.


Formally, a complete frequency plan is defined as the assignment, for every beam $b \in \{1, ..., N_B\}$, of the following elements:
\begin{itemize}
    \item A discrete number of bandwidth slots $b_b$, which can't be greater than $N_{BW}$.
    \item A positive integer $f_b$, that indicates the first bandwidth slot used. Beam $b$ then uses slots $f_b, f_b + 1, ..., f_b + b_b - 1$, all of them part of the available spectrum. 
    \item A positive integer $r_b$ representing a frequency reuse out of the $N_{FR}$ available.
    \item In case $N_P=2$, a binary variable representing the chosen polarization.
\end{itemize}

\begin{figure*}[!t]
\centering
\includegraphics[width=.8\linewidth]{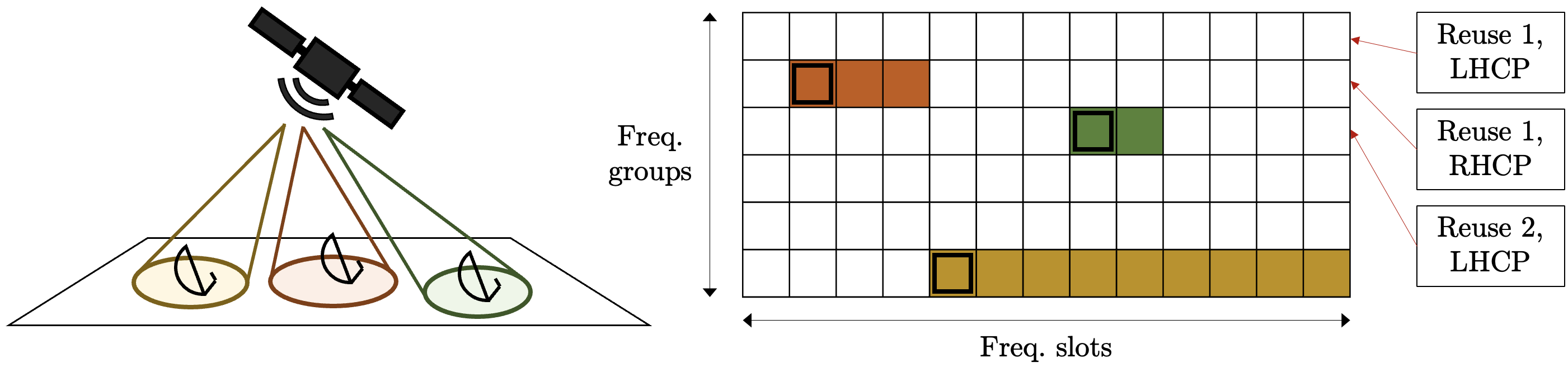}
\caption{Frequency assignment representation in the form of a grid with $N_{FR}\cdot N_P$ rows and $N_{BW}$ columns. In this example, $N_{FR}=3$, $N_P=2$, and $N_{BW}=13$. Each of the 3 beams being powered by the satellite is assigned to a cell in the grid representing the first slot (black squares) and to a certain number of consecutive slots (colored cells). For instance, the beam depicted in green is assigned to reuse group 2, left polarization, and is using slots 8 and 9.}
\label{fig:freqplan}
\end{figure*}

Figure \ref{fig:freqplan} introduces a representation of this decision space in the form of a grid, with $N_{FR}\cdot N_P$ rows and $N_{BW}$ columns. Each column represents a frequency slot, whereas each row corresponds to a combination of a frequency reuse and a polarization. As shown in the figure, rows are sorted, first, by frequency reuse, and second, by polarization. With this representation, making a frequency assignment for a beam turns into picking a specific cell in the grid corresponding to the first slot (black squares in the figure) and choosing a valid number of slots (colored cells).

\begin{figure*}[!t]
\centering
\includegraphics[width=.84\linewidth]{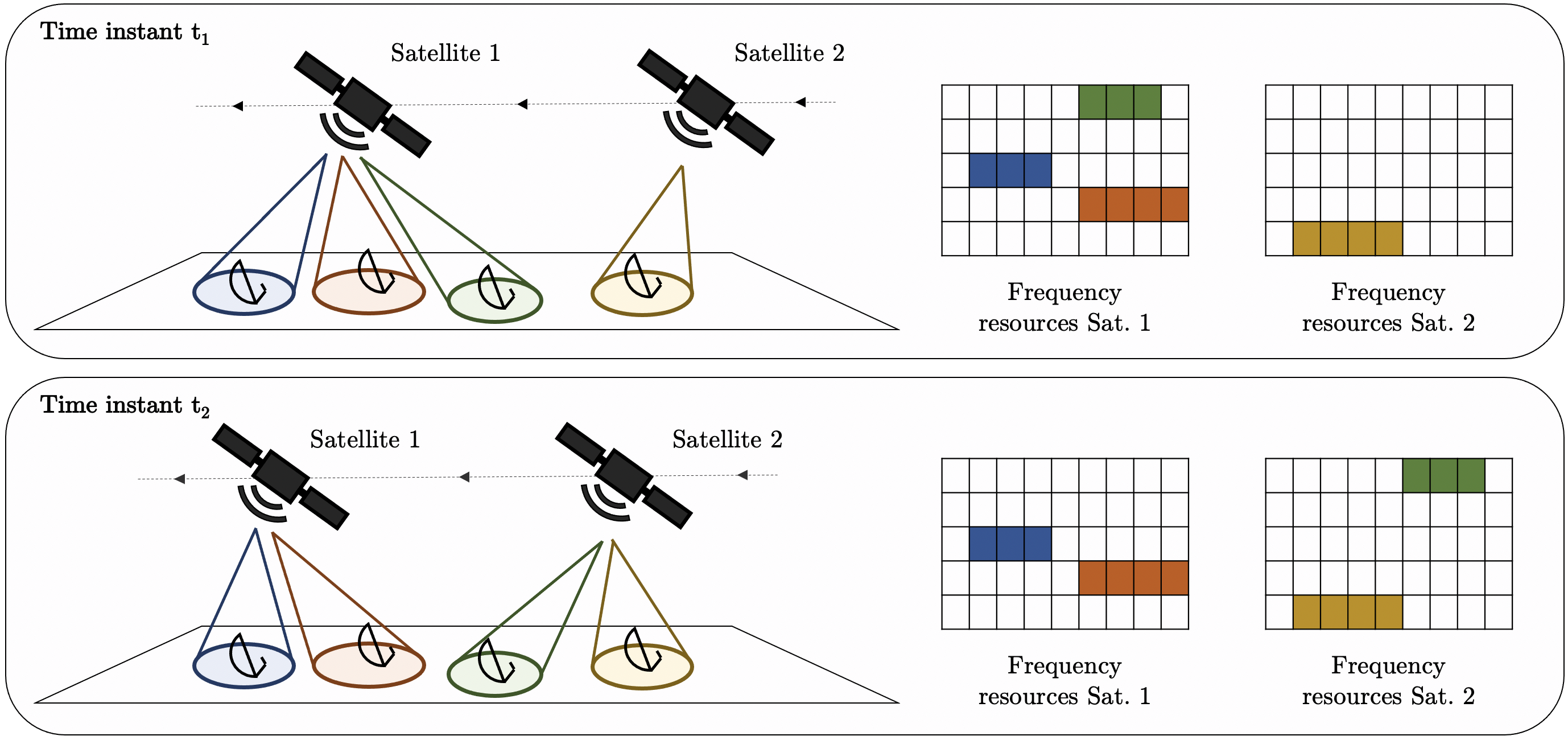}
\caption{Handover operation example between two satellites at time instants $t_1$ and $t_2$.}
\label{fig:handover}
\end{figure*}

As previously described, based on the upcoming satellite communications landscape, we assume all satellites have digital and flexible payloads. Therefore, during operations, we assume any of the assignment parameters can be changed in real-time at negligible cost. Generally though, the operator will choose to make changes to the frequency assignment of certain beams whenever those beams undergo a handover. If the frequency assignment of a beam is not changed during the handover, then the beam utilizes the same frequency assignment on the new satellite. This is depicted in Figure \ref{fig:handover}, in which the green beam switches from satellite 1 to satellite 2 and preserves the same frequency reuse, polarization, and bandwidth slots. 

When a handover occurs, it is critical that the resources to be used on the new satellite are not being used by any other beam already. This constitutes an important constraint when making frequency assignment decisions for a NGSO constellation and applies to any pair of beams being powered by the same satellite at any point in time. We define this constraint for a pair of beams as an \emph{intra-group} restriction between those beams. Then, the set $\mathcal{R}_A$ represents all pairs $(i, j)$ such that beams $i$ and $j$ hold an intra-group restriction. We assume that this set is known in advance and is externally updated during operation if needed.

\begin{figure}[!t]
\centering
\includegraphics[width=.5\linewidth]{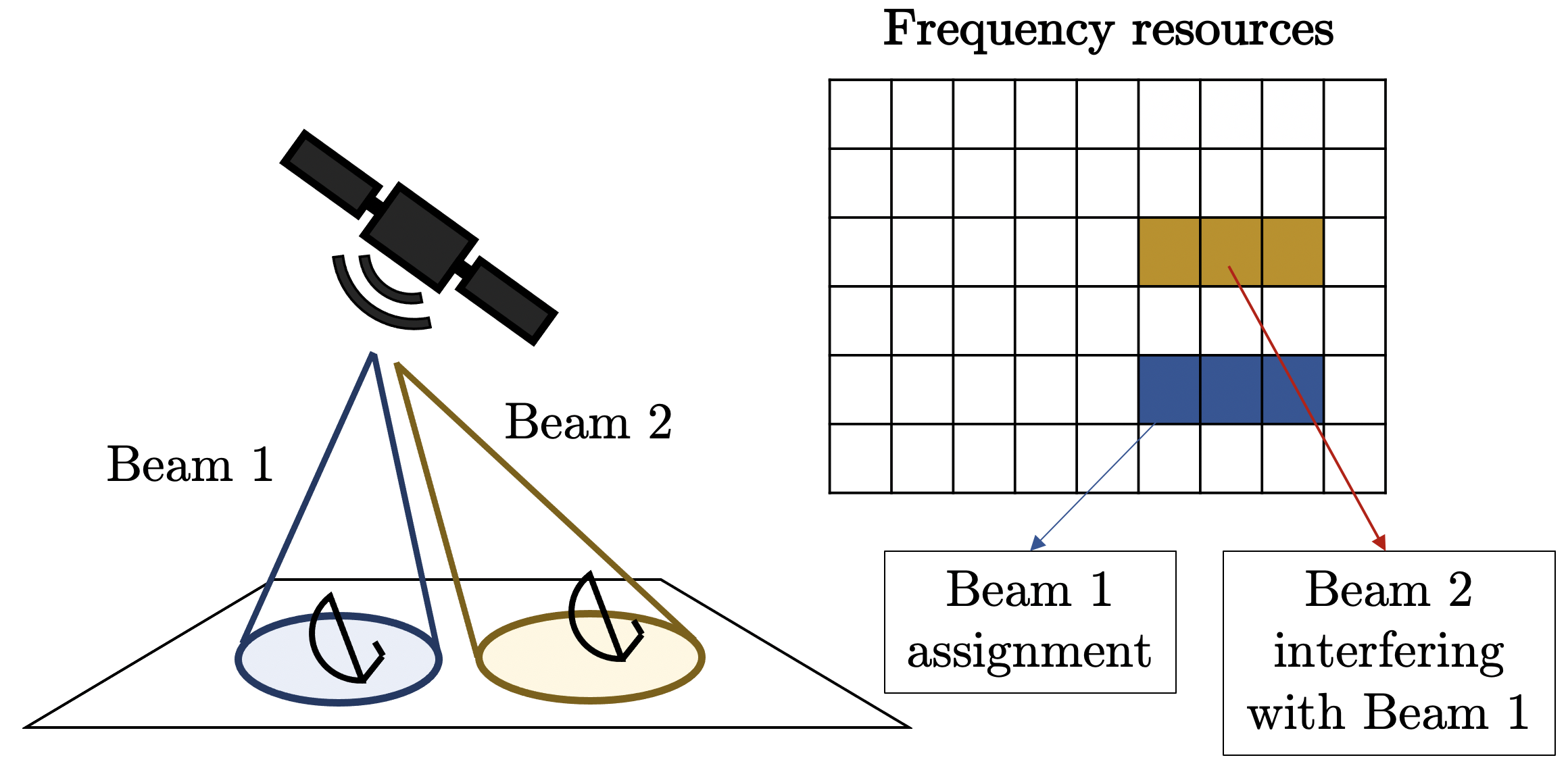}
\caption{Inter-group restriction example between beam 1 and beam 2. Diagram shows the moment beam 2 is to be assigned and beam 1 has already been assigned.}
\label{fig:inter}
\end{figure}

There is another type of constraint to consider. If two beams whose footprints are close use the same polarization, they might interfere with each other if their assigned frequency slots overlap. This event, which we define as an \emph{inter-group} restriction, is represented in Figure \ref{fig:inter}. Similarly to the handover constraint case, the set $\mathcal{R}_E$ encodes all pairs of beams $(i, j)$ that hold an inter-group restriction. This set is constructed following some interference criteria (e.g., SINR level, angular distance threshold) and might be time-dependant if the constellation serves mobile users.

\section{Proposed Optimization Method}
\label{sec:opt_framework}

In this section we encode each decision and restriction presented in the previous section as variables and constraints of an optimization method based on Integer Linear Programming (ILP). In the context of satellite resource allocation, methods such as ILP allow operators to easily encode problem needs, objectives, and constraints with low granularity with respect to the decision variables. Then, there exist multiple commercially-available tools that automatically produce the solution of each program. 

In each subsection we outline how a specific feature of the frequency assignment problem is encoded in the ILP formulation using linear operators. We also give an overview of possible objective functions to guide the search to select the best feasible frequency plan. We assume the constraint sets $\mathcal{R}_A$ and $\mathcal{R}_E$ are an input to the model.

\subsection{Frequency Plan Decisions}

On the beam level, the frequency assignment decisions consist of choosing how many bandwidth slots, which ones, and which frequency reuse and polarization to use. From the perspective of the grid representation introduced in Figure \ref{fig:freqplan}, this means selecting 1) a column and 2) a row in the grid, and then 3) a number of consecutive slots. We encode the column (i.e., the first slot) as an integer variable $f_i$, with domain $\{1, N_{BW}\}$, for each beam $i\in\{1,...,N_B\}$. Then we encode the row (i.e., frequency reuse and polarization) as an integer variable $g_i$, with domain $\{1,...,N_{FR}\cdot N_P\}$. Finally, the number of consecutive slots is encoded as an integer variable $b_i$, with the same domain as $f_i$. We introduce these variables in the ILP formulation as follows: 
\begin{alignat}{2}
    \label{eq:fi}
    & f_i \in \{1,..., N_{BW}\},\enspace&\forall\,i\in\{1,...,N_B\} \\ \label{eq:gi}
    & g_i \in \{1,..., N_{FR}\cdot N_P\},\enspace&\forall\,i\in\{1,...,N_B\} \\
    \label{eq:bi}
    & b_i \in \{1,..., N_{BW}\},\enspace&\forall\,i\in\{1,...,N_B\}
\end{alignat}
Any frequency plan can then be decoded using these three variables per beam. The optimization method returns these values to the operator. 

\subsection{Constraint: Limited spectrum}

We now start accounting for the constraints of the problem. We first encode the constraint that all bandwidth slots used must be within the limits imposed by the system (i.e., only the $N_{BW}$ considered slots can be used). This is encoded as follows:
\begin{alignat}{2}
    \label{eq:limited_spectrum}
    f_i + b_i - 1 \leq N_{BW},\enspace\forall i\in\{1,...,N_B\}
\end{alignat}

\subsection{Auxiliary variables: relative spectrum location}

The remaining constraints account for the intra-group and inter-group restrictions. Simply put, both of these restrictions involve the need to avoid spectrum overlaps for certain pairs of beams. For instance, if beams $i$ and $j$ hold a restriction, and are assigned to the same frequency reuse and polarization (i.e., $g_i = g_j$), we want to make sure that $f_i + b_i \leq f_j$ or $f_j + b_j \leq f_i$, i.e., we want to ensure that beam $i$ has allocated spectrum either to the left or to the right of beam $j$, but without overlapping. Before diving into the actual restrictions, these two possibilities need to be encoded in the optimization formulation in order to preserve linearity. To that end, we use binary variables $z_{ij}$ and add the following constraints:
\begin{alignat}{2}
    & z_{ij}\in\{0,1\},\enspace& \forall\,i,j\,\,\text{s.t.}\,(i,j)\in\mathcal{R} \\ \label{eq:z1}
    & f_j - f_i \geq 0 - M(1 - z_{ij}),\enspace& \forall\,i,j\,\,\text{s.t.}\,(i,j)\in\mathcal{R} \\ \label{eq:z0}
    & f_i - f_j\geq\epsilon - Mz_{ij},\enspace&\forall\,i,j\,\,\text{s.t.}\,(i,j)\in\mathcal{R}
\end{alignat}
where $M$ is a ``sufficiently large'' number, $\epsilon$ is a very small positive number, and $\mathcal{R}$ represents the union $\mathcal{R}_A\cup\mathcal{R}_E$. Given a restriction $(i,j)$, if $z_{ij} = 1$, (\ref{eq:z1}) is active and enforces that $f_j \geq f_i$ (i.e., beam $j$ can not use lower frequencies than beam $i$'s). On the contrary, if $z_{ij} = 0$, (\ref{eq:z0}) is active and the effect is the opposite.

\subsection{Constraint: Intra-group or handover restrictions}

We start by considering the intra-group restrictions, given by the set $\mathcal{R}_A$. This type of restrictions are caused by handover operations and are relevant if and only if constrained beams use the same frequency reuse and polarization. To first encode this condition linearly, we define auxiliary binary variables $y_{ij}$ alongside the following constraints:
\begin{alignat}{2}
    \label{eq:y0}
    & y_{ij}\in\{0,1\},\enspace& \forall\,i,j\,\,\text{s.t.}\,(i,j)\in\mathcal{R}_A \\ \label{eq:y1}
    & g_i \geq g_j - M(1-y_{ij}),\enspace&\forall\,i,j\,\,\text{s.t.}\,(i,j)\in\mathcal{R}_A\\ \label{eq:y2}
    & g_i \leq g_j + M(1-y_{ij}),\enspace&\forall\,i,j\,\,\text{s.t.}\,(i,j)\in\mathcal{R}_A
\end{alignat}
If $y_{ij} = 1$, the method enforces that $g_i = g_j$, since both (\ref{eq:y1}) and (\ref{eq:y2}) are active. If $y_{ij} = 0$, the opposite should occur; however, this can not be achieved solely with variable $y_{ij}$. To enforce strict inequality when $y_{ij} = 0$, we introduce binary variables $p_{ij}$ to account for both cases $g_i > g_j$ and $g_j > g_i$. Hence, we add the following constraints:
\begin{alignat}{2}
    \label{eq:pp}
    & p_{ij}\in\{0,1\},\enspace&\forall\,i,j\,\,\text{s.t.}\,(i,j)\in\mathcal{R}_A \\ \label{eq:p1}
    & g_i - g_j\geq\epsilon -M(1-p_{ij} + y_{ij}),\enspace&\forall\,i,j\,\,\text{s.t.}\,(i,j)\in\mathcal{R}_A \\ \label{eq:p0}
    & g_i - g_j \leq -\epsilon + M(p_{ij}+y_{ij}),\enspace&\forall\,i,j\,\,\text{s.t.}\,(i,j)\in\mathcal{R}_A
\end{alignat}
which are active if and only if $y_{ij}=0$. In that case, and when $p_{ij} = 1$, (\ref{eq:p1}) is active and we have $g_i > g_j$. On the contrary, if $p_{ij} = 0$, (\ref{eq:p0}) is active and $g_j > g_i$ holds. To summarize the effect of the auxiliary variables introduced so far, Table \ref{tab:auxiliary_intra} shows how the different frequency assignment cases between two beams holding an intra-group restriction are encoded by means of variables $z_{ij}$, $y_{ij}$, and $p_{ij}$.

\begin{table}[t]
\renewcommand{\arraystretch}{1.15}
\centering
\caption{Encoding of different frequency assignment cases by means of auxiliary variables when beams $i$ and $j$ share an \textit{intra-group} constraint.}
\label{tab:auxiliary_intra}
\begin{tabular}{c|c}
\hline
\rowcolor[HTML]{EFEFEF} 
\textbf{Auxiliary variables}  & \multicolumn{1}{c}{\cellcolor[HTML]{EFEFEF}\textbf{Frequency assignment case}} \\ \hline
$z_{ij} = 1$  &  $f_i \leq f_j$ \\
$z_{ij} = 0$  &  $f_i > f_j$ \\
\hline
$y_{ij} = 1$ &  $g_i = g_j$ (same freq. reuse and polarization) \\
$y_{ij} = 0$ and $p_{ij} = 1$  &  $g_i > g_j$ \\
$y_{ij} = 0$ and $p_{ij} = 0$  &  $g_i < g_j$ \\
\hline
\end{tabular}
\end{table}

Finally we introduce the constraints that encode the intra-group restrictions:
\begin{alignat}{2}
    \label{eq:intra1}
    & f_i + b_i\leq f_j + M(2-y_{ij}-z_{ij}), \enspace&\forall\,i,j\,\,\text{s.t.}\,(i,j)\in\mathcal{R}_A \\ \label{eq:intra2}
    & f_j + b_j\leq f_i + M(1-y_{ij}+z_{ij}), \enspace&\forall\,i,j\,\,\text{s.t.}\,(i,j)\in\mathcal{R}_A
\end{alignat}
where $M$ is again a ``sufficiently large'' number. These constraints enforce a non-overlapping frequency assignment between beams $i$ and $j$ holding an intra-group restriction if and only if both beams use the same reuse group and polarization ($y_{ij} = 1$). Constraint (\ref{eq:intra1}) does so for the case in which $z_{ij}=1$ ($f_i \leq f_j$), whereas constraint (\ref{eq:intra2}) is active when $z_{ij}=0$ ($f_i > f_j$).

\subsection{Constraint: Inter-group or interference restrictions}

The inter-group restrictions are given by set $\mathcal{R}_E$ and concern all pairs of beams with close footprints, which might interfere with each other during operations. Setting a threshold for how close two interfering beams can be is up to the operator's policy, which might prefer to trade additional inter-group restrictions for further interference mitigation. One way to define this set might be to impose an angular distance threshold between beams; we leave this decision out of the scope of this paper. 

As introduced in Figure \ref{fig:inter}, this type of restrictions can negatively impact the performance of the system when both beams are using the same polarization, regardless of their frequency reuse. To specifically focus on polarization, we first introduce the following variables and constraints:
\begin{alignat}{2}
    & k_i \in \{1,...,N_{FR}\},\enspace&\forall\,i\in\{1,...,N_B\} \\
    & m_i \in \{0, N_P -1\},\enspace&\forall\,i\in\{1,...,N_B\} \\
    & g_i = N_Pk_i - m_i,\enspace&\forall\,i\in\{1,...,N_B\}
\end{alignat}
Variable $k_i$ encodes the frequency reuse assigned to beam $i$ whereas variable $m_i$ encodes its polarization, if any.

Similar to constraints (\ref{eq:y0}) - (\ref{eq:y2}), we introduce binary variable $s_{ij}$ for each pair of beams holding an inter-group constraint. This variable encodes whether beams $i$ and $j$ use the same polarization, by means of the following constraints:
\begin{alignat}{2}
    & s_{ij} \in \{0,N_P-1\},\enspace&\forall\,i,j\,\,\text{s.t.}\,(i,j)\in\mathcal{R}_E \\ \label{eq:s1}
    & m_i \geq m_j - Ms_{ij},\enspace&\forall\,i,j\,\,\text{s.t.}\,(i,j)\in\mathcal{R}_E \\ \label{eq:s2}
    & m_i \leq m_j + Ms_{ij},\enspace&\forall\,i,j\,\,\text{s.t.}\,(i,j)\in\mathcal{R}_E
\end{alignat}
If $s_{ij} = 0$ then both (\ref{eq:s1}) and (\ref{eq:s2}) are active, and $m_i$ and $m_j$ are enforced to be equal. Note that in case $N_P=1$, then $s_{ij}$ is always zero, since there is only one polarization.

Then, following the same idea behind constraints (\ref{eq:pp}) - (\ref{eq:p0}), we use binary variable $d_{ij}$ to help encoding whether $m_i > m_j$ or $m_j > m_i$, i.e., enforcing $m_i$ and $m_j$ to be different in case there is more than one polarization and $s_{ij}=1$. The following constraints explain this idea:
\begin{alignat}{2}
    & d_{ij} \in\{0,1\},\enspace&\forall\,i,j\,\,\text{s.t.}\,(i,j)\in\mathcal{R}_E \\
    \label{eq:d0}
    & m_i - m_j \leq -\epsilon + M(1 + d_{ij} - s_{ij}),\enspace&\forall\,i,j\,\,\text{s.t.}\,(i,j)\in\mathcal{R}_E \\
    \label{eq:d1}
    & m_i - m_j \geq \epsilon - M(2-d_{ij}+ s_{ij}),\enspace&\forall\,i,j\,\,\text{s.t.}\,(i,j)\in\mathcal{R}_E
\end{alignat}
which can only be active if $s_{ij} = 1$ (more than one polarization). Then, if $d_{ij} = 0$, (\ref{eq:d0}) is active and therefore $m_j > m_i$. On the other hand, if $d_{ij}=1$, (\ref{eq:d1}) is active and $m_i > m_j$. Table \ref{tab:auxiliary_inter} summarizes all cases that inter-group restrictions-related auxiliary variables encode.
\begin{table}[t]
\renewcommand{\arraystretch}{1.15}
\centering
\caption{Encoding of different frequency assignment cases by means of auxiliary variables when beams $i$ and $j$ share an \textit{inter-group} constraint.}
\label{tab:auxiliary_inter}
\begin{tabular}{c|c}
\hline
\rowcolor[HTML]{EFEFEF} 
\textbf{Auxiliary variables}  & \multicolumn{1}{c}{\cellcolor[HTML]{EFEFEF}\textbf{Frequency assignment case}} \\ \hline
$z_{ij} = 1$  &  $f_i \leq f_j$ \\
$z_{ij} = 0$  &  $f_i > f_j$ \\
\hline
$s_{ij} = 0$ &  $m_i = m_j$ (same polarization) \\
$s_{ij} = 1$ and $d_{ij} = 0$  &  $m_i < m_j$ \\
$s_{ij} = 1$ and $d_{ij} = 1$  &  $m_i > m_j$ \\
\hline
\end{tabular}
\end{table}

Lastly, the actual constraints encoding the inter-group restriction interactions are the following:
\begin{alignat}{2}
    \label{eq:inter1}
    & f_i + b_i \leq f_j + M(1 + s_{ij} - z_{ij}), \enspace&\forall\,i,j\,\,\text{s.t.}\,(i,j)\in\mathcal{R}_E \\
    \label{eq:inter2}
    & f_j + b_j \leq f_i + M(s_{ij} + z_{ij}), \enspace&\forall\,i,j\,\,\text{s.t.}\,(i,j)\in\mathcal{R}_E
\end{alignat}
These constraints enforce a non-overlapping assignment between beams $i$ and $j$ holding an inter-group restriction if and only if both beams use the same polarization ($s_{ij}=0$). Constraint (\ref{eq:inter1}) does so for the case in which $z_{ij}=1$ ($f_j \geq f_i$), whereas constraint (\ref{eq:inter2}) is active when $z_{ij}=0$ ($f_i > f_j$).

\subsection{Objective Function}

So far, we have discussed the decisions and constraints that need to be encoded to define \emph{valid} frequency plans, i.e., plans that do not violate any constraint. However, we have not addressed how to prioritize different valid frequency plans according to the operator's preferences and goals. In certain occasions, operators might prefer plans that maximize bandwidth allocation or plans that use as few frequency reuses as possible. To encode these preferences into the ILP formulation, we propose the following objective function to be maximized:
\begin{equation}
\label{eq:obj_function}
    \underset{}{\text{max}}\,\,\sum_{i=1}^{N_B}\left(\beta_{1,i}b_i - |\beta_{2,i}|g_i - |\beta_{3,i}|f_i -|\beta_{4,i}|P_i(f_i, b_i)\right)
\end{equation}
where $\beta_{1,i}$, $\beta_{2,i}$, $\beta_{3,i}$, and $\beta_{4,i}$ are weighting parameters for beam $i$, with $\beta_{k,i}\in\mathbb{R}$. This function combines four different objectives:
\begin{enumerate}
    \item $\beta_{1,i}b_i$ focuses on maximizing allocated bandwidth for beam $i$, or minimizing it in case $\beta_{1,i} < 0$. A better control over used bandwidth can lead to a more efficient power consumption.
    \item $-|\beta_{2,i}|g_i$ attempts to use as few frequency reuses as possible, which is the case if $|\beta_{2,i}| > 0$. Setting $\beta_{2,i} = 0$ for all beams enables a uniform use of frequency reuses and polarizations. The coefficient $\beta_{2,i}$ uses absolute terms because the effects of maximizing and minimizing $g_i$ are symmetric according to our formulation.
    \item $-|\beta_{3,i}|f_i$ seeks to have the same effect on the number of frequency slots used. If $-|\beta_{3,i}|<0$, lower parts of the spectrum are prioritized. If $\beta_{3,i}=0$, then the spectrum is used uniformly. We specifically subtract this term given lower frequencies may require less power consumption, and therefore are sometimes preferred by operators.
    \item If available, $-|\beta_{4,i}|P_i(f_i, b_i)$ directly represents the RF power consumed by beam $i$ when using slots $f_i, f_i+1, ..., f_i+b_i-1$. Reducing RF power when possible is reflected by this operand when $\beta_{4,i}>0$. In our experiments, this metric does not correspond to a linear combination of variables per se. In the following section we provide additional details on its computation.
\end{enumerate}
The weighting parameters are used to define a priority hierarchy over these objectives. While these parameters can be identical for all beams, operators might be interested in prioritizing additional bandwidth or certain bands for specific beams, for example.

\subsection{Additional design constraints}

The variables and constraints presented so far encode the necessary information connected to the main decisions and main restrictions of the problem. There are other type of considerations that are not critical to operation or might not always be applicable but might be part of the operator's desiderata. We describe them next.

\subsubsection{Minimum bandwidth requirement}

We have defined the bandwidth variable (\ref{eq:bi}) with domain $\{1,...,N_{BW}\}$. However, we might be interested in specifying higher lower bounds, for contractual reasons or since using a single bandwidth slot might not be enough to satisfy the link budget equation \cite{maral2011satellite} for certain beams. This way, we could redefine variable (\ref{eq:bi}) as
\begin{alignat}{2}
    & b_i \in \{c_i,..., N_{BW}\},\enspace&\forall\,i\in\{1,...,N_B\} \tag{\ref{eq:bi}}
\end{alignat}
where $c_i$ is the minimum number of slots that beam $i$ requires.




\subsubsection{Business constraints}

While not a requirement of the problem per se, operators might prefer to first split frequency resources between different sets of users. For instance, specific frequency reuses might be saved for mobile users or the spectrum might be partitioned according to user latitudes. These constraints can be easily encoded in the ILP formulation by modifying the corresponding domains of variables (\ref{eq:gi}) and (\ref{eq:bi}).

\subsubsection{Active and inactive beams}

In this work we have assumed a feasible solution is always possible. However, we might find ourselves in the unlikely scenario in which the operator must turn off certain beams in order to achieve constraint satisfaction. In this case, which beams to turn off and how to redesign the frequency assignment constitutes a different optimization problem that needs additional variables. Specifically, variables $a_i$ are introduced to encode whether beam $i$ should be active or inactive/turned off in the plan. We describe the remaining changes with respect to the presented ILP formulation in Appendix \ref{sec:active}.

\section{Results}
\label{sec:results}
In this section we present and discuss the results and performance of the proposed frequency plan optimization method. Specifically, we carry out three different experiments to validate its usefulness: 1) first, we test our algorithm on three low-dimensional scenarios in which spectrum usage maximization is required. 2) In the second experiment, we study the scalability of our method by optimizing the frequency assignment in larger systems. To that end, we introduce an iteration-based optimization procedure that allows us to decrease the overall optimization time when needed. Finally, 3) in the third experiment, we consider a multibeam satellite constellation based on a real large constellation and a realistic user distribution to assess the benefits of using our approach to reduce on-board power consumption.

\begin{table}[ht]
\renewcommand{\arraystretch}{1.15}
\centering
\caption{Dimensionality of the satellite system and frequency assignment parameters used in each experiment. *$N_{ch}$ corresponds to the number of changes per iteration when the iteration-based adaptation is used (see Section \ref{sec:exp2}). In all cases we consider a seven-satellite constellation ($N_S = 7$) and two polarizations ($N_P=2$).}
\label{tab:scenarios}
\begin{tabular}{c*{5}{|c}}
\hline
\rowcolor[HTML]{EFEFEF} 
Experiment  & Freq. reuses $N_{FR}$ & Bandwidth slots $N_{BW}$ & Users $N_U$  & Beams $N_B$ & Changes $N_{ch}$*\\ \hline
1 & 8 & 40 & 50, 60, 100 & 96, 118, 182 & - \\
2 & 20 & 200 & 1,000 & 1,060 & 25, 50, 100\\
3 & 20 & 200 & 20,000 & 5,000 & 50 \\
\hline
\end{tabular}
\end{table}

In all cases, the experiments make use of a representative example of a MEO constellation system, the O3b mPOWER constellation filed by O3b limited \cite{SES}. In this case, it has $N_S=7$ satellites, each capable of using $N_P=2$ polarizations. User distributions as well as gateway locations are provided by SES S.A. based on realistic configurations. In our analyses, we focus on optimizing the downlink frequency assignment, and vary the number of beams $N_B$, frequency reuses $N_{FR}$, and bandwidth slots $N_{BW}$ to represent scenarios with different spectrum availability. The specific parameters considered in each of the three experiments are summarized in Table \ref{tab:scenarios}. The variable $N_{ch}$ corresponds to a feature of the iteration-based procedure described in Section \ref{sec:exp2}. We use a commercial solver, Gurobi, for our experiments \cite{gurobi}.

Before the method begins the optimization, given a particular user distribution, we use the beam placement algorithm described in \cite{PachlerdelaOsa2020} to determine the required number of beams and their positions. We also make use of the heuristic frequency assignment algorithm described in the same reference as a baseline benchmark throughout the remainder of the paper. In addition, this algorithm serves as a warm-start when using the iteration-based procedure detailed in Section \ref{sec:exp2}. Based on this beam placement, we consider two beams will have an inter-group or interference restriction if the center of their footprints is closer than four times the half-cone angle.

\subsection{Experiment 1: Maximizing Bandwidth Allocation}
\label{exp1}

In this experiment, our goal is to verify the behavior of the optimization method by maximizing bandwidth allocation in three scenarios with an increasing number of users and, in turn, an increasing number of beams, as described in Table \ref{tab:scenarios}. We encode bandwidth allocation maximization in the proposed objective function (\ref{eq:obj_function}) by adjusting the weighting parameters as follows:
$$
    \beta_{1,i} = 1,\enspace\beta_{2,i}=\beta_{3,i}=\beta_{4,i}=0, \quad \forall i\in\{1,...,N_B\}
$$
To quantify the performance of our method and compare it against others, we use the normalized total assigned bandwidth for both the heuristic ($BW_{HEU}$) and optimized ($BW_{ILP}$) frequency plans. This metric is defined as follows:
\begin{equation}
    BW = \frac{1}{C_{tot}}\sum_{i=1}^{N_B}b_{i}
\end{equation}
where $C_{tot}$ is the total system capacity, defined as the total number of frequency slots available in the constellation:
\begin{equation}
    C_{tot} = N_S\cdot N_{BW}\cdot N_{FR} \cdot N_P
\end{equation}
As mentioned, the rationale behind improving spectrum utilization is that it generally results in lower power consumption, without explicitly resorting to the introduction of power considerations in the formulation by means of parameters $\beta_{4, i}$. We explore this case in Section \ref{exp3}. While we do not restrict the maximum bandwidth that can be assigned per beam, in real operations this might be necessary due to spectral efficiency considerations \cite{maral2011satellite}. If needed, a bandwidth upper bound can be also encoded in our formulation.


\begin{figure}[!t]
\centering
\subfloat[Baseline freq. plan \cite{PachlerdelaOsa2020}]{%
    \includegraphics[width=0.25\linewidth]{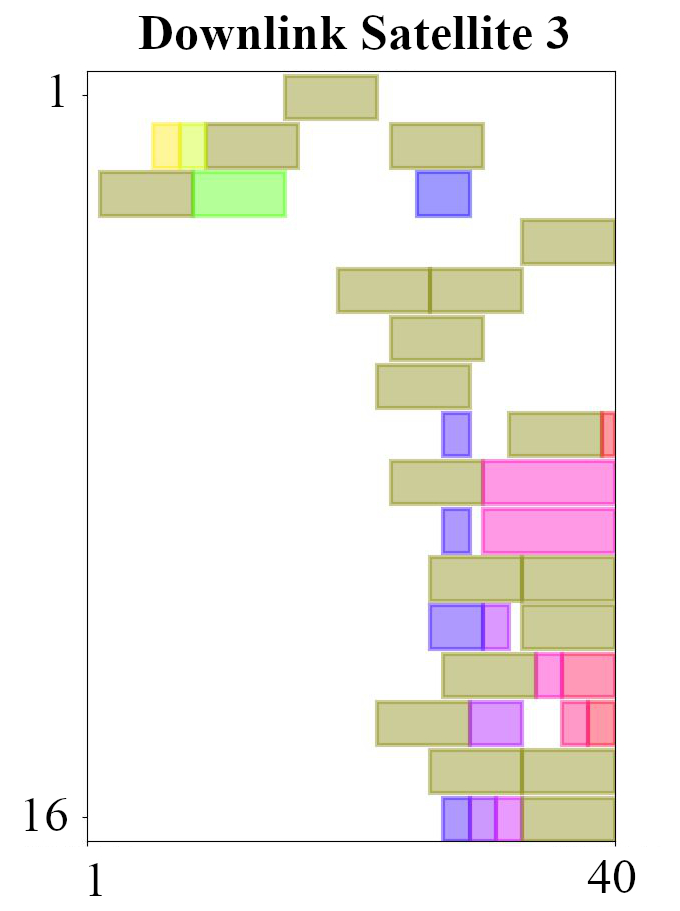}}
    \qquad
\subfloat[Optimized freq. plan (ILP)]{%
    \includegraphics[width=0.25\linewidth]{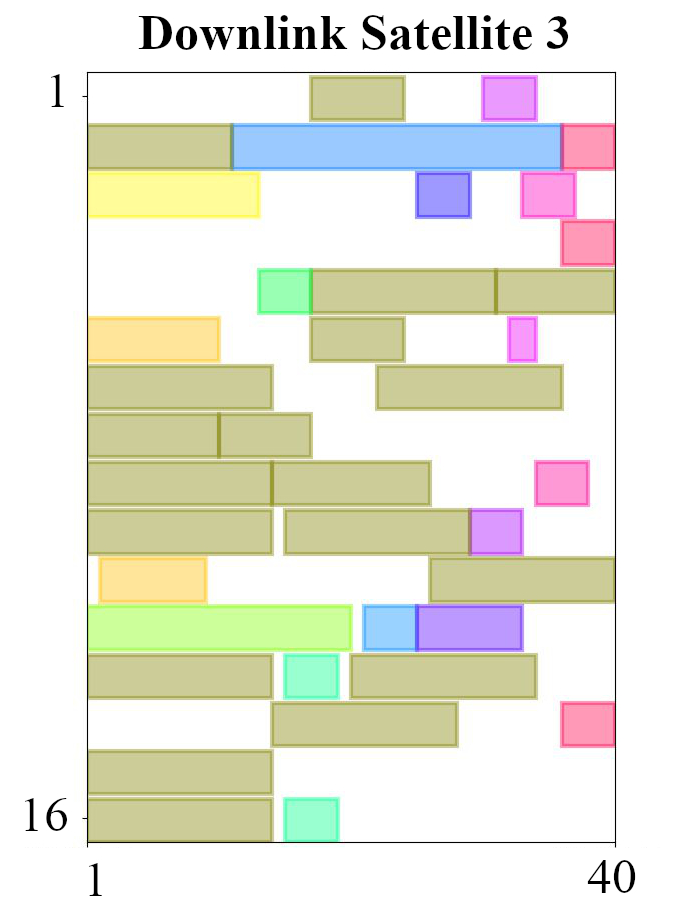}}
\caption{Comparison for one specific satellite in the constellation between the frequency assignment produced by the heuristic algorithm serving as baseline (left) and our method (right) when bandwidth maximization is prioritized. The scenario is defined with $N_S=7$, $N_P=2$, $N_{FR}=8$, $N_{BW}=40$, and $N_B=182$. The horizontal axis represents the frequency slots $f_i\in\{1,...,N_{BW}\}$, whereas the vertical axis represents the frequency reuse and polarizations $g_i\in\{1,...,N_{FR}\cdot N_P\}$. Gold coloring is used for gateway beams, while a different hue is used depending on the central frequency assigned to each user beam.}
\hfill
\label{fig:exp1_frequ_plan}
\end{figure}

Figure \ref{fig:exp1_frequ_plan} shows the frequency assignment at a specific instant for one of the satellites when using the heuristic algorithm and when using our method. Each colored region represents the frequency slots assigned to a beam. As observed, the colored region of the plot is substantially larger when using our method. Since the plot represents the frequency assignment for a specific time instant corresponding to a specific routing of beams to satellites, unassigned slots might be reserved for beams that are soon to undergo or have recently undergone handover operations in the simulation.


To quantify the increase in the assigned bandwidth, Table \ref{tab:exp1_results} presents the normalized total assigned bandwidth for the heuristic algorithm and our method in the three test scenarios. The results show that our optimization method achieves up to a 73\% bandwidth increase with respect to the heuristic baseline solution in the scenario with 100 users, while satisfying all frequency and handover constraints. 

An increase in the number of beams might not be proportional to an increase in the total number of frequency slots that we are able to allocate. This is clearly reflected by the values of $BW_{ILP}$ presented in Table \ref{tab:exp1_results}, since moving from 96 beams to 118 beams entails a total assigned bandwidth increase, but the effect of moving from 118 beams to 182 beams is the opposite. In the latter case, when increasing the number of beams, the number of intra- and inter-group constraints also increases, and therefore frequency assigments can not exploit more frequency slots in order to avoid frequency overlaps. Still, our method makes the best assignment possible regardless of the number of constraints.

\begin{table}[!t]
    \centering
    \caption{Results for Experiment 1, focused on maximizing bandwidth allocation.}
    \label{tab:exp1_results}
    \begin{tabular}{c*{5}{|c}}
        \hline
        \rowcolor[HTML]{EFEFEF}
        Scenario & Users $N_U$ & Beams $N_B$ & $BW_{HEU}$ & $BW_{ILP}$ & $BW$ Increase \\ \hline
        1.1 & 50 & 96 & 0.18 & 0.27 & 51\% \\
        1.2 & 60 & 118 & 0.25 & 0.41 & 61\% \\
        1.3 & 100 & 182 & 0.21 & 0.37 & 73\% \\
        \hline
    \end{tabular}
\end{table}



\subsection{Experiment 2: Iteration-based extension and computing time tradeoffs}
\label{sec:exp2}



In this experiment we test the scalability of our method and develop an iteration-based procedure to reduce the computing time in high-dimensional scenarios. As the dimensionality of both the constellation and user base increase, making frequency decisions in real time becomes more challenging, since the number of constraints scales quadratically with the number of beams. If computing time is a constraint, there is a point in which every algorithm---even state-of-the-art ones---must make tradeoffs in order to meet time requirements. We explore the tradeoffs of our method in this section.

A method to speed up the optimization when dimensionality is large is presented in Appendix \ref{sec:iteration}. In this modified approach, we transform the prior formulation to an iteration-based approach where the search space is restricted to interesting regions. For each beam, we select a few variable assignments for each decision variable and let the solver choose the best combinations across beams. In each iteration, the optimization algorithm changes the frequency assignment of $N_{ch}$ beams, while the rest are kept fixed. Although the algorithm can make frequency decisions without any prior input, we found that using a complete suboptimal frequency plan as a warm-start substantially decreases runtime. There is no requirement of validity for the warm-start frequency plan, since our method can amend violated constraints or deactivate beams. In our experiments, we use the frequency plan provided by the heuristic algorithm presented in \cite{PachlerdelaOsa2020} as a warm-start. In general, we achieve a significant reduction in the number of variables and constraints that are considered in each call to the optimizer.


To test the iterative optimization method, we use a scenario with higher dimensionality, with $N_{FR}=20$ frequency reuses, $N_{BW}=200$ bandwidth slots, and $N_U=1,000$ users. The users and their corresponding gateways are connected using $N_B=1,060$ beams. We therefore increase both the search space ---more resources--- and the constraint set ---more beams--- compared to the first experiments. We still optimize the frequency plan to maximize the allocated bandwidth. We run the algorithm using three different configurations: $N_{ch}=25$, $50$, and $100$ changes at each iteration. The beams assigned at every iteration are chosen randomly and we do not impose any restriction on how many times a single beam can undergo assignment. Given the nature of the procedure, we might require specific beams to be reassigned in different iterations before converging to a stable solution. In our experiments we run the algorithm until getting this stable solution; we consider it is reached when the total allocated bandwidth does not increase during 50 consecutive iterations. Since we introduce stochastic elements, we run the algorithm 10 times for each of the selected $N_{ch}$ and report statistics on the 10 runs.

\begin{figure}[!t]
    \centering
    \includegraphics[width=.65\linewidth]{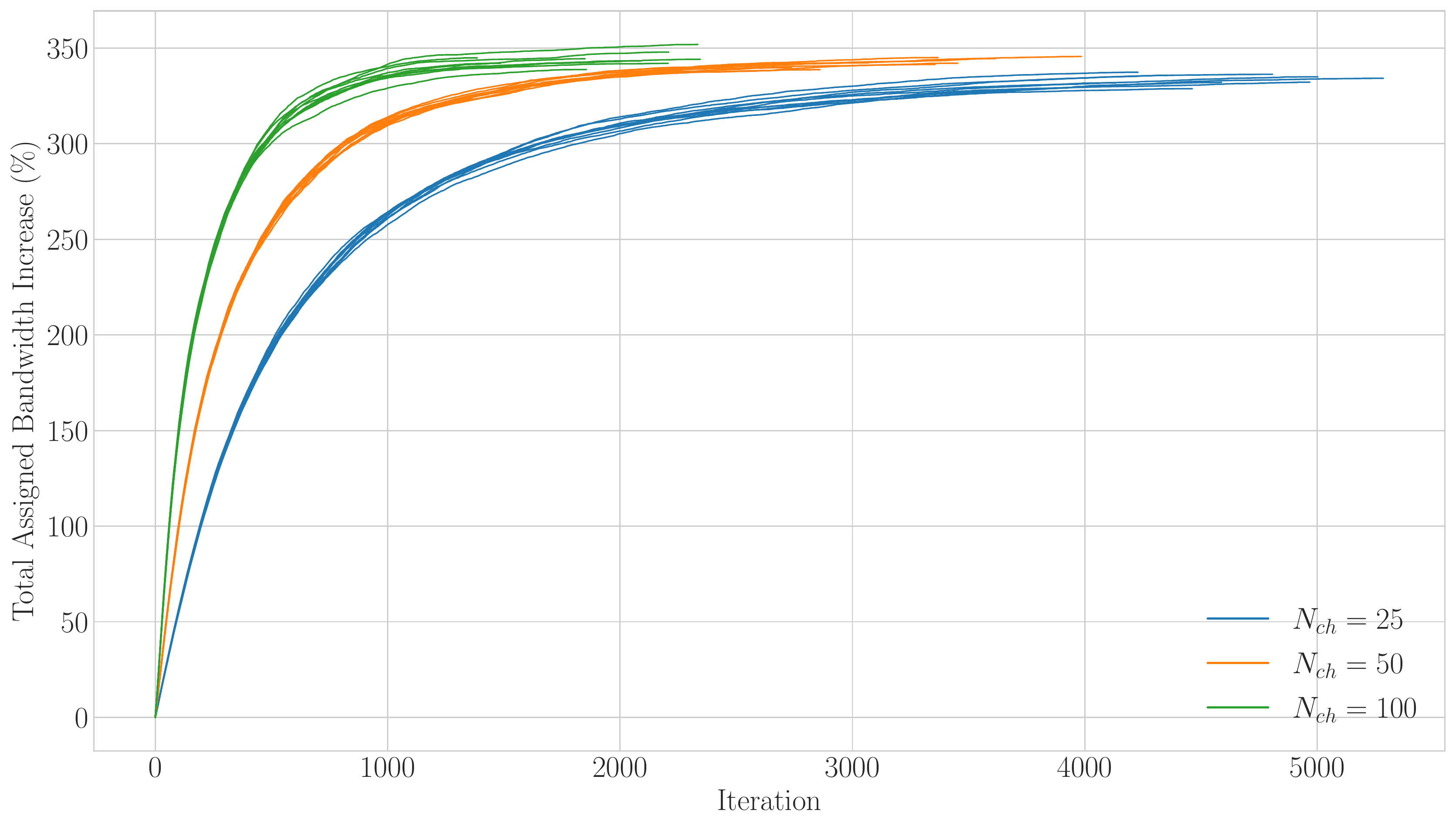}
    \caption{Evolution of the percentual increase of the total allocated bandwidth at each iteration for different values of $N_{ch}$.}
    \label{fig:exp2_sum_bw_it}
\end{figure}

Table \ref{tab:exp2_results} presents the results of using the iterative approach to reach a stable solution in less time when using $N_{ch}=25$, $50$, and $100$. We compare the total assigned bandwidth increase with respect to the warm-start and the computing time (wall-clock time). The value of $N_{ch}$ is directly related to the tradeoff between computing time and number of iterations. In all cases, we are able to find a global frequency assignment that more than triples the total assigned bandwidth without violating any constraints of the system. When using $N_{ch}=25$ we do so in less than one third the amount of time required for $N_{ch}=100$, although, since we are assigning fewer beams at a time, we observe that the improvement per iteration is lower, as graphically represented in Figure \ref{fig:exp2_sum_bw_it}. The average total number of assignments, i.e. $N_{ch}\cdot N_{it}$, is approximately $1.1\times10^5$, $1.6\times10^5$, and $2.0\times10^5$ for $N_{ch}=25$, $50$, and $100$, respectively. Note that this is about two orders of magnitude higher than the number of beams ($N_B=1,060$ in these experiments). Table \ref{tab:exp2_results_breakdown} provides additional insights on the time and iterations it takes to increase the bandwidth by 100\%, 200\%, and 300\%. Generally, we can observe our method achieves 80\% of the improvement in less than 20\% of the convergence time.

\begin{table*}[!t]
    \centering
     \caption{Results for Experiment 2, reported over 10 different runs and focused on maximizing bandwidth allocation using the iterative optimization method. All simulations run on a server with 20 cores of an Intel 8160 processor and 192 GB of RAM. SEM = Standard Error of the Mean.}
    \label{tab:exp2_results}
    \begin{tabular}{c|c*{4}{|D{,}{\pm}{-1}}}
\hline
\rowcolor[HTML]{EFEFEF}
 & 
 & \multicolumn{1}{c|}{$BW_{ILP}$}
 & \multicolumn{1}{c|}{$BW$ Increase (\%)}
 & \multicolumn{1}{c|}{Iterations $N_{it}$}
 & \multicolumn{1}{c}{Comp. time (min)}
 \\
\rowcolor[HTML]{EFEFEF}
\multirow{-2}{*}{Changes $N_{ch}$}
 & \multirow{-2}{*}{$BW_{HEU}$}
 & \text{Mean},\text{SEM}
 & \text{Mean},\text{SEM}
 & \text{Mean},\text{SEM}
 & \text{Mean},\text{SEM}
 \\
\hline
25  & \multicolumn{1}{c|}{\multirow{3}{*}{0.14}} & 0.61,0.00 & 333,1 & 4578,130 & 10.4,0.2 \\
50 & & 0.63,0.00 & 342,1 & 3188,138 & 18.9,0.8 \\
100 & & 0.63,0.00 & 344,1 & 1967,101 & 46.3,1.8 \\
\hline
    \end{tabular}
\end{table*}

\begin{table*}[!t]
    \centering
     \caption{Results for Experiment 2, number of iterations and the computing time required to reach an increase in allocated bandwidth of 100\%, 200\%, and 300\% for three different algorithm configurations: $N_{ch}=25$, $50$, and $100$. Number of iterations and computing time required to converge are also included. All simulations run on a server with 20 cores of an Intel 8160 processor and 192 GB of RAM. SEM = Standard Error of the Mean.}
    \label{tab:exp2_results_breakdown}
    \begin{tabular}{c*{8}{|D{,}{\pm}{-1}}}
\hline
\rowcolor[HTML]{EFEFEF}
 & \multicolumn{4}{c|}{Iterations $N_{it}$}
 & \multicolumn{4}{c}{Comp. time (min)}
 \\
\rowcolor[HTML]{EFEFEF}
\multirow{-2}{*}{Changes $N_{ch}$}
 & \multicolumn{4}{c|}{$\text{Mean}\pm\text{SEM}$}
 & \multicolumn{4}{c}{$\text{Mean}\pm\text{SEM}$}
 \\
\hline
\rowcolor[HTML]{E0E0E0}
$BW$ Increase
 & \multicolumn{1}{c|}{100\%}
 & \multicolumn{1}{c|}{200\%}
 & \multicolumn{1}{c|}{300\%}
 & \multicolumn{1}{c|}{Convergence} 
 & \multicolumn{1}{c|}{100\%}
 & \multicolumn{1}{c|}{200\%}
 & \multicolumn{1}{c|}{300\%}
 & \multicolumn{1}{c}{Convergence} \\
\hline
25 & 195,1 & 534,3 & 1673,21 & 4578,130 & 2.3,0.7 & 3.6,1.1 & 5.9,1.9 & 10.4,3.3 \\
50 & 101,0 & 281,1 & 831,7 & 3188,138 & 2.9,0.9 & 4.8,1.5 & 8.1,2.6 & 18.9,6.0 \\
100 & 61,0 & 165,1 & 469,5 & 1967,101 & 6.0,1.9 & 10.1,3.2 & 18.0,5.7 & 46.3,14.6 \\
\hline
    \end{tabular}
\end{table*}



With these experiments, we assess that the iteration-based procedure enables the scalability of our optimization method, which is an important requirement for future high-dimensional constellation operations. Although limiting the number of beams that the optimization algorithms can modify significantly reduces the dimensionality of the solution space, the performance improvement comes at the expense of converging to suboptimal solutions and, therefore, needing to assign around one hundred times more beams in total. Nonetheless, robustness and fast reaction times are generally preferred over optimality when operating in highly dynamic and time-dependent environments, where the feasibility of the frequency assignment might be only temporary. Appropriately setting hyperparameters such as $N_{ch}$ is key in those cases \cite{aliaga22a}.

\subsection{Experiment 3: Minimizing Power Consumption in a Real-World Scenario}\label{exp3}

\begin{figure*}[!t]
\centering
\hfill
\subfloat[Baseline frequency plan used as warm-start (computed using algorithm from \cite{PachlerdelaOsa2020}) ]{%
    \includegraphics[width=1.0\linewidth]{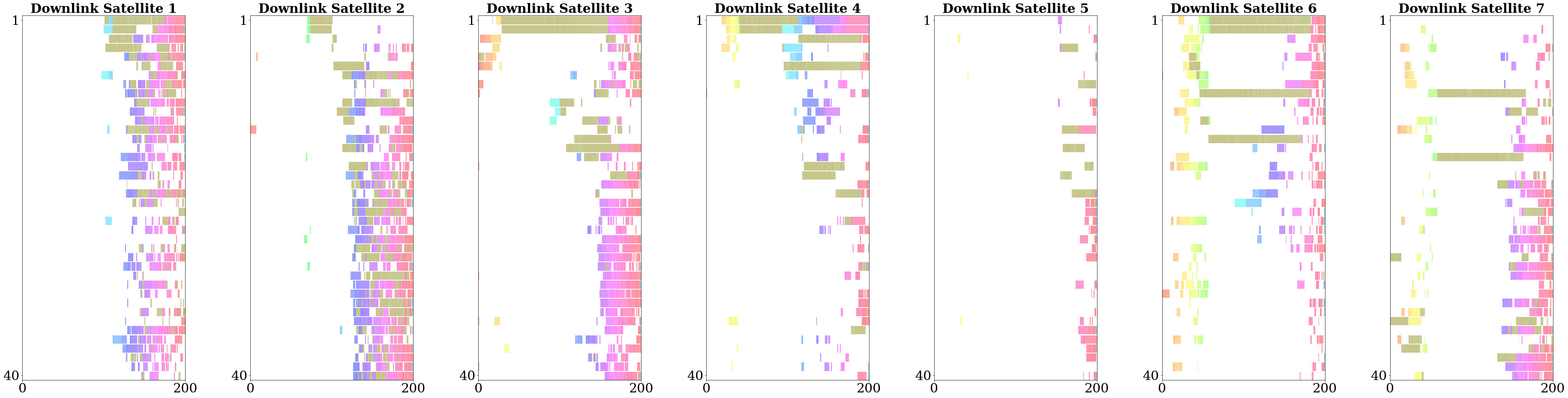}}
\\
\subfloat[Optimized frequency plan (our method)]{%
    \includegraphics[width=1.0\linewidth]{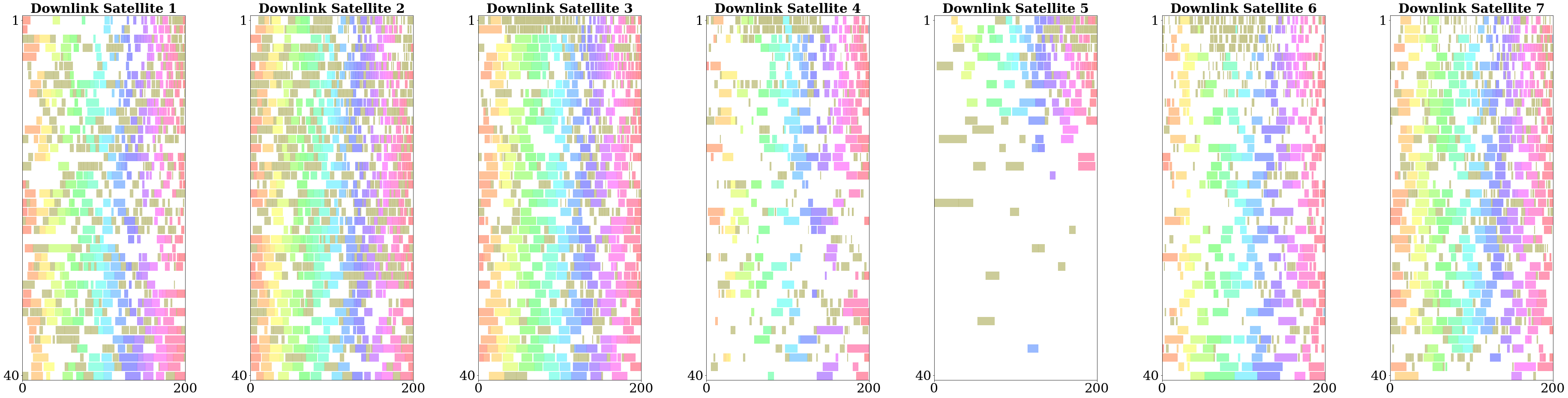}}
\caption{Comparison between the warm-start and the optimized frequency assignment in a scenario with $N_S=7$, $N_P=2$, $N_{FR}=20$, $N_{BW}=200$, and $N_B=5000$.}
\hfill
\label{fig:exp3_frequency_plan}
\end{figure*}

Finally, in this third experiment, we demonstrate the benefits of using our method to optimize the frequency assignment in a realistic scenario with a large-scale constellation and user base. The user data is provided by SES S.A. and includes throughput demand for 20,000 users located around the globe which, after running our beam placement algorithm, are grouped into approximately 5,000 beams (see Appendix \ref{sec:user_distribution}). In this experiment we prioritize directly reducing power consumption by leveraging the fourth term of the objective function equation (\ref{eq:obj_function}), $-|\beta_{1,i}|P_i(f_i, b_i)$. For the purposes of this section, we understand $P_i$ as a linear function that returns the consumed power for beam $i$ given its $f_i$ and $b_i$ assignments. Its full implementation details into the formulation are described in Appendix \ref{sec:power_consumption}. 

\begin{table}[!t]
    \centering
    \caption{Results for Experiment 3, minimizing power consumption in a real-world scenario.}
    \label{tab:exp3_results}
    \begin{tabular}{c*{5}{|c}}
        \hline
\rowcolor[HTML]{EFEFEF}

 Changes $N_{ch}$
 & Power Decrease
 & $BW_{HEU}$
 & $BW_{ILP}$
 & $BW$ Increase
 & Iterations $N_{it}$
 \\
\hline
50 
& 39.8\%
& 0.175 & 0.614 & 251\%
& 2464
\\
\hline
    \end{tabular}
\end{table}



Figure \ref{fig:exp3_frequency_plan} shows the warm-start (heuristic baseline algorithm) and the optimized frequency plan (our method) for the seven satellites of the constellation at one moment of the simulation. For this experiment, we run the iterative implementation of the optimization algorithm with $N_{ch}=50$ until a stable solution is reached, as it can provide a solution under real-time operational constraints. The results presented in Table \ref{tab:exp3_results} show that using our method reduces the power consumption by up to a 40\% after fewer than 2,500 iterations. This is achieved by increasing the total allocated bandwidth by more than 250\%, which can be observed in Figure \ref{fig:exp3_frequency_plan}: the optimized plan has significantly larger colored regions, i.e., assigned frequency slots, as well as several unused combinations of a frequency reuse and polarization.

The results of this experiment demonstrate that the framework with the iteration-based procedure is able to substantially improve the frequency assignment in a high-dimensional, real-world-based system. The algorithm is able to navigate the search space of this highly-constrained problem and change the frequency allocation of beams to decrease the overall power consumption. The number of changes per iteration $N_{ch}$ and the total number of iterations can be modified to adjust the system's operation to meet any possible time requirements. Although we used a specific representative constellation for our experiments, our framework can be adapted to constellations with different flexibilities: number of satellites and planes, number of frequency reuses and slots, additional operational constraints, etc.

\section{Conclusion}
\label{sec:conclusions}
In this paper, we have addressed the need to develop a frequency assignment optimization algorithm that accounts for the flexibility and dimensionality of the upcoming communication satellite operations. Our optimization framework leverages the use of frequency reuse mechanisms to efficiently handle carrier and bandwidth assignment for multisatellite constellations with thousands of beams while respecting handover and frequency constraints, a set of requirements that has previously not been addressed altogether. Our method is based on Integer Linear Programming and produces optimal or quasi-optimal frequency plans according to a flexible objective function that allows encoding multiple goals such as maximizing bandwidth, minimizing the number of active frequency reuses, and minimizing RF power consumption. We have also presented an iteration-based implementation of the framework that not only enables its operation in high-dimensional use cases but also introduces degrees of flexibility to configure it depending on the scenario and computing time constraints. 

We have carried out three different experiments to validate the performance, scalability, and potential real-world operability of our method, respectively. The results of the first experiment prove the ability of the framework to efficiently solve the frequency plan design problem, achieving a 73\% improvement on the total allocated bandwidth in simple scenarios. The iterative optimization method analyzed in the second experiment is able to optimize the frequency assignment in higher-dimensional scenarios, achieving over a 300\% improvement on spectrum usage in a time-efficient manner. Finally, the results from the third experiment demonstrate the benefits of using our framework in a real-world scenario with approximately 20,000 users and 5,000 beams. In this case, the framework increases the allocated bandwidth but prioritizes making an efficient use of frequency reuses and polarizations in order to achieve a 40\% reduction in on-board power consumption.



%

\appendices
\section{Optimization with Activation Variables}
\label{sec:active}

It is possible that, given the constraints of the problem, not all conditions can be met and hence feasibility can not be guaranteed. To minimize the impact in those situations, some beams can be turned off and their assigned resources reallocated. We introduce binary variables $a_i$ to encode whether beam $i$ should be active or not:
\begin{equation}
    a_i\in\{0,1\},\,\enspace\forall\,i\in\{1,\dots,N_B\}
\end{equation}
If beam $i$ is active, then $a_i = 1$. Now these variables need to be taken into consideration when specifying the rest of the constraints of the program. Below we describe the changes to the constraints presented in the original framework.

\subsection{Limited spectrum}

If the maximum number of bandwidth slots available are not enough for beam $i$ to satisfy the link budget equations \cite{maral2011satellite}, the beam should be turned off. To encode this scenario we modify constraint (\ref{eq:limited_spectrum}) and add the activation variables:
\begin{alignat}{2}
    f_i + b_i - 1 \leq N_{BW} + M(1 - a_i),\enspace\forall i\in\{1,...,N_B\} \tag{\ref{eq:limited_spectrum}}
\end{alignat}
Here, variable $M$ represents a ``sufficiently large'' number, such that this constraint has no effect if beam $i$ is not active ($a_i = 0$).

\subsection{Intra-group restrictions}

Other reasons why we might need to turn off certain beams are related to intra-group and inter-group restrictions. Regarding the former, constraints (\ref{eq:intra1}) and (\ref{eq:intra2}) can be extended to account for beam activation:
\begin{alignat}{2}
    & f_i + b_i - M(1-a_i)\leq f_j + M(3-a_j-y_{ij}-z_{ij}),\enspace&\forall\,i,j\,\,\text{s.t.}\,(i,j)\in\mathcal{R}_A \tag{\ref{eq:intra1}} \\
    & f_j + b_j - M(1-a_j)\leq f_i + M(2-a_i-y_{ij}+z_{ij}),\enspace&\forall\,i,j\,\,\text{s.t.}\,(i,j)\in\mathcal{R}_A \tag{\ref{eq:intra2}}
\end{alignat}
Now these constraints prevent frequency overlap between beams $i$ and $j$ holding an intra-group restriction if and only if both beams are active ($a_i=1$ and $a_j=1$).

\subsection{Inter-group restrictions}

Similarly, the constraints encoding the inter-group restrictions (\ref{eq:inter1}) and (\ref{eq:inter2}) are also extended:
\begin{alignat}{2}
    & f_i + b_i - M(1-a_i) \leq f_j +  M(2 - a_j + s_{ij} - z_{ij}),\enspace&\forall\,i,j\,\,\text{s.t.}\,(i,j)\in\mathcal{R}_E \tag{\ref{eq:inter1}} \\
    & f_j + b_j - M(1 - a_j) \leq f_i + M(1 - a_i + s_{ij} + z_{ij}),\enspace&\forall\,i,j\,\,\text{s.t.}\,(i,j)\in\mathcal{R}_E \tag{\ref{eq:inter2}}
\end{alignat}

\subsection{Objective function}

Finally, to prevent the framework from unnecessarily turning off beams to improve the objective function (\ref{eq:obj_function}), we include an additional term in the function that accounts for beam activation:
\begin{equation}
    \underset{}{\text{max}}\,\,\sum_{i=1}^{N_B}\left(\beta_{1,i}b_i -  |\beta_{2,i}|g_i - |\beta_{3,i}|f_i - |\beta_{4,i}|P_i(f_i, b_i) + |\beta_{5,i}|a_i\right) \tag{\ref{eq:obj_function}}
\end{equation}
By increasing the weight of parameter $\beta_{5,i} \in \mathbb{R}$ the framework will prioritize having beam $i$ active to maximizing any of the other elements of the function.


\section{Iteration-based optimization formulation}
\label{sec:iteration}

There are two main factors that influence the computing time of an ILP formulation: the number of decisions to make and the dimensionality of the search space. An efficient formulation is able to reduce both of those factors without compromising the quality of the solution. The following lines describe the improved iteration-based ILP formulation used in Sections \ref{sec:exp2} and \ref{exp3}. Note that, contrary to the initial formulation, the speed-ups proposed do not guarantee convergence or optimality over the final solution, but substantially reduce the amount of computing time needed to obtain a \textit{good enough} result.

\subsection{From full-optimization to an iteration-based approach}
The first element to notice is that optimizing a large number of beams at the same time entails a high-dimensional combinatorial problem. Without accounting for interference restrictions, each beam has an order of $N_{FR}N_{P}N_{BW}^{2}$ possibilities: for each beam, we can decide the frequency reuse, polarization, initial frequency, and total bandwidth to use. Then, the number of frequency plan combinations is in the order of $(N_{FR}N_{P}N_{BW}^{2})^{N_{B}}$. While this number is still tractable for low dimensional scenarios, optimizing for a large number of beams is infeasible from a time perspective. A way to reduce the dimensionality of the problem is to tune only certain beams at a time, while keeping the rest fixed, and iterate over the optimization procedure while selecting different sets of beams so that all beams are optimized. The number of beams that are allowed to change at every iteration will be denoted by $N_{ch}$.

\subsection{Reducing the search space by ranking}
Next, we note that, although we can decide from $N_{FR}N_{P}N_{BW}^{2}$ possibilities for each beam, only a small subset of those options is interesting from an optimization point of view; we explain this below. Specifically, considering the activation variables introduced in Appendix \ref{sec:active}, each beam $i$ contributes to the objective function in the form of:
\begin{equation}
    \beta_{1,i}b_i - |\beta_{2,i}|g_i - |\beta_{3,i}|f_i -|\beta_{4,i}|P_i(f_i, b_i) + |\beta_{5,i}|a_i \tag{\ref{eq:obj_function}}
\end{equation}
This allows us to rank the different options based on their contribution to the total objective value. The most interesting options will be the ones with the highests ranks. Let us denote $x_{f,g,b,i}$ as a binary decision variable that represents choosing the option for beam $i$ which has initial frequency $f$, frequency reuse $g$, and bandwidth $b$. We can rewrite the contribution of each beam as:
\begin{equation}
    \sum_{f,g,b}(\beta_{1,i}b - |\beta_{2,i}|g - |\beta_{3,i}|f -|\beta_{4,i}|P_i(f, b)+|\beta_{5,i}|)x_{f,g,b,i}=\sum_{f,g,b}l_{f,g,b,i}x_{f,g,b,i}
\end{equation}
Where $l_{f,g,b,i}$ is the contribution of the option denoted by $x_{f,g,b,i}$. Note that now the last factor $|\beta_{5,i}|$ does not include the activation variable $a_{i}$. This is addressed in the following section. Now, instead of considering all the feasible options, we will consider only a subset of alternatives for each beam, denoted by the symbol $\mathcal{V}_i$. By considering a strict subset, we are limiting the options of the algorithm and reducing the search space of solutions, with the expectation that the optimal or close to optimal solution lies within the selected subset. By including a ranking system, we can define this subset as the $X$ most interesting options. It is important to highlight that, while solutions with higher bandwidth will tend to be preferred, those are also the most difficult to allocate in the frequency plan. Therefore, instead of a single ranking system for all options, we pre-compute a ranking for each possible bandwidth, and include the options in the top of each ranking in the set. In our experiments we select the top 10 candidates per bandwidth assignment. Also, note that the interference constraints (inter-group restrictions) with beams that are not being changed can be computed \textit{a priori}, and options that are not feasible (i.e., solutions that have interference constraints with beams that are not being changed), do not need to be included in the ranking, thus reducing the number of active constraints in the problem. The new problem formulation can be stated as:
\begin{equation}
    \underset{}{\text{max}}\,\,\sum_{i=1}^{N_B}\sum_{f,g,b \in \mathcal{V}_{i}}l_{f,g,b,i}x_{f,g,b,i}
\end{equation}
Additionally, we need to impose that at most one option can be active at a time:
\begin{equation}
    \label{eq:iteration_based_one_option_at_a_time}
    \sum_{f,g,b \in \mathcal{V}_{i}}x_{f,g,b,i} = 1 \ \forall i
\end{equation}

\subsection{Dealing with activation variables}

Now we explain how the activation variables can come into play in this new formulation. As we use a warm-start in our iteration-based experiments, we first need to distinguish between cases where the beam in the warm-start plan is active or not:
\begin{enumerate}
    \item According to the warm-start, the original beam $i$ is active, and it has initial frequency $f_{orig}$, frequency reuse $g_{orig}$, and bandwidth $b_{orig}$. Then, we include an additional option $x_{orig,i}$ that corresponds to the equivalent option in the original plan. Then, the previous constraint can be expressed as:
    \begin{equation}
        \sum_{f,g,b \in \mathcal{V}_{i}}x_{f,g,b,i} + x_{orig,i} = 1 \tag{\ref{eq:iteration_based_one_option_at_a_time}}
    \end{equation}
    \item According to the warm-start, the original beam $i$ is not active. Then, we include the additional variable $a_{i}$ that indicates if the beam is active ($a_{i}=1$) or not ($a_{i}=0$). The previous constraint can be expressed as:
    \begin{equation}
        \sum_{f,g,b \in \mathcal{V}_{i}}x_{f,g,b,i} - a_{i} = 0 \tag{\ref{eq:iteration_based_one_option_at_a_time}}
    \end{equation}
\end{enumerate}
With this formulation, a feasible option is guaranteed to exist, although not all beams might be active in the final solution. If a feasible solution with all beams active does exist, this formulation will provide it.

\subsection{Including intra- and inter-group restrictions}
The final aspect to address in this new approach is how to include intra- and inter-group constraints. As mentioned, interference with fixed beams is passively included by disregarding the infeasible options. The only restrictions to be included are the ones between beams that are allowed to change in the same iteration. Since the decision variables are now just binary activation variables with implicit frequency reuse, initial frequency, and bandwidth, we can easily precompute if two options from two different beams collide or not. If they do, we only need to add a restriction of the type:
\begin{equation}
    x_{f_{i},g_{i},b_{i},i} + x_{f_{j},g_{j},b_{j},j} \leq 1
\end{equation}
Which ensures that at most one of the two options will be active, guaranteeing constraint satisfaction in the final solution. Note that, since we need to add a constraint for each pair of colliding options, the number of constraints may increase quadratically with the number of options, which imposes computational constraints in the number of options that can be considered.

Now, the complexity of the problem is solely determined by the number of beams that are allowed to change at each point, and the number of options considered for each beam, which are engineering decisions that can be assessed independently of the problem characteristics. As a final remark, note that if we set $N_{ch}=N_{B}$ and consider all valid options, the formulation is equivalent to the original formulation presented in this work.

\section{Beam placement distribution}
\label{sec:user_distribution}

\begin{figure}[htbp]
    \centering
    \includegraphics[width=\linewidth]{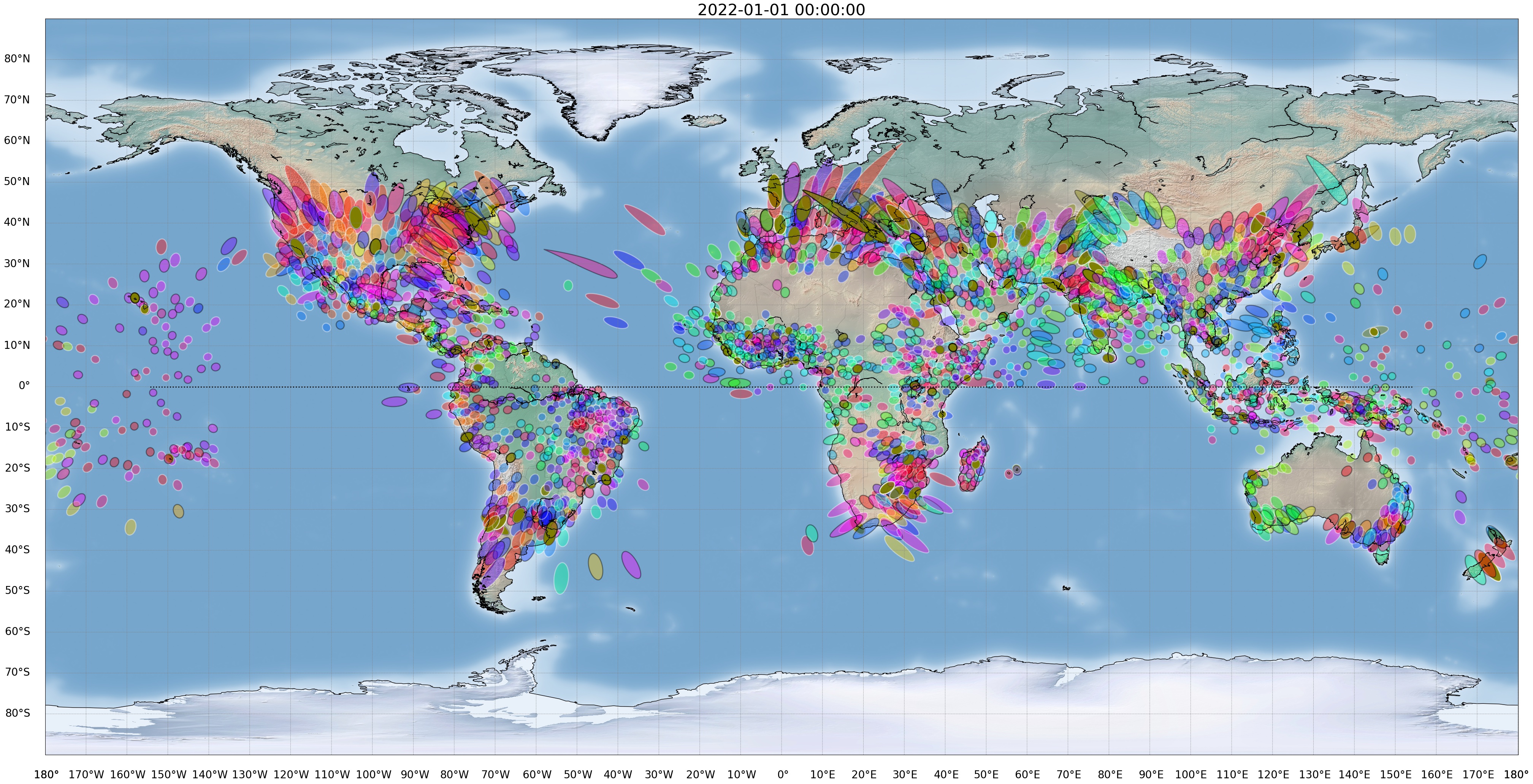}
    \caption{Beam placement for scenario with approximately 18,000 users and 5,000 beams.}
    \label{fig:exp3_distribution}
\end{figure}

Figure \ref{fig:exp3_distribution} shows the beam placement distribution that results after using the beam placement algorithm from \cite{PachlerdelaOsa2020} on a set of approximately 18,000 users. 5,000 beams are used to serve these users.

\section{Power computing algorithm}
\label{sec:power_consumption}

One of the elements in the objective function equation (\ref{eq:obj_function}) corresponds to a metric for power consumption. As power equations are not linear \cite{maral2011satellite}, we can not directly use them in our framework. As a solution, we precompute, for each of the $N_B$ beams, the consumed power for each possible frequency assignment.

To that end, we use the satellite communications models described in \cite{paris19}. For simplicity, we describe the procedure to compute the necessary power $P_b$ for one beam $b$ given its data rate demand $D_b$ and a certain allocated bandwidth $BW_b$. We assume the satellites use the MODCOD schemes defined in the standards DVB-S2 and DVB-S2X \cite{DVB2015}. Given a certain roll-off factor $\alpha_b$, we can compute the lower bound of the required spectral efficiency as
\begin{equation}
    \Gamma_{req} = \frac{D_b(1 + \alpha_b)}{BW_b}
\end{equation}
We select the MODCOD whose spectral efficiency is the lowest such that $\Gamma \geq \Gamma_{req}$. If no such MODCOD exists, we set a power value of $P_b = M$, where $M$ is a sufficiently large number.

Otherwise, we can get the appropriate value for $E_b/N$ from the MODCOD scheme. Since in our work we have considered interference mitigation mechanisms by means of the inter-group constraints, we assume interference is negligible. We then compute the necessary $C/N_0$ as
\begin{equation}
    \left.\frac{C}{N_0}\right|_b = \left.\frac{E_b}{N}\right|_b \cdot \frac{D_b}{BW_b}
\end{equation}
With $C/N_0$ in dB, we can then compute the power as:
\begin{align}
    P_b =& \, \left.\frac{C}{N_0}\right|_b + OBO - G_{T_x} - G_{R_x}  \nonumber\\ & + \text{FSPL} + 10\log_{10}(kT_{sys})
\end{align}
where OBO is the power-amplifier output back-off, $G_{T_x}$ and $G_{R_x}$ are the transmitting and receiving antenna gains, respectively, $k$. is the Boltzmann constant, and $T_{sys}$ is the system temperature (assumed to be 290K). FSPL and $L_{atm}$ account for the free-space path losses, respectively. We assume FSPL are significantly larger than atmospheric losses, and losses at the transmitting and receiving antennas, so we neglect all the latter.

We compute a power value $P_b$ for each possible assignment of $BW_b$, and repeat the process for all beams in the constellation.

\section*{Acknowledgment}
This work was supported by SES S.A.. The authors would like to thank SES S.A. for their input to this paper and their financial support. The project that produced these results also received the support of a fellowship from “la Caixa” Foundation (ID 100010434). The fellowship code is LCF/BQ/AA19/11720036.

\ifCLASSOPTIONcaptionsoff
  \newpage
\fi



%


\bibliographystyle{IEEEtran}
\bibliography{library}
%








\end{document}